\documentclass[11pt,twoside]{article}

\usepackage{asp2006}
\usepackage{epsf}
\usepackage{epsfig}
\usepackage{lscape}

\begin{document}
\title{Eddington limited starbursts in the central 10\,pc of AGN, and
  the Torus in NGC\,1068}
\author{R. Davies, R. Genzel, L. Tacconi, F. Mueller S\'anchez}
\affil{Max Planck Institut f\"ur extraterrestrische Physik, Garching, Germany}
\author{A. Sternberg}
\affil{School of Physics \& Astronomy, Tel Aviv University, Tel Aviv, Israel}

\begin{abstract}
We present results from a survey of nearby AGN using the near infrared
adaptive optics integral field spectrograph SINFONI.
These data enable us to probe the distribution and kinematics of the
gas and stars at spatial resolutions as small as 0.085$''$.
We find strong evidence for recent but short lived starbursts residing
in very dense nuclear disks;
on scales of less than 10\,pc these would have reached
Eddington-limited luminosities when active, perhaps accounting for
their short duration. 
In addition, for NGC\,1068 at a resolution of 6\,pc, we present direct
observations of molecular gas close around the AGN which we identify
with the obscuring torus.
\end{abstract}


\section{Introduction}
\label{dav:sec:intro}

We have mapped the distribution and
kinematics of the gas and stars in 9 nearby AGN using adaptive optics
to reach high spatial resolution at near infrared wavelengths.
The AGN are mostly Seyfert~1s, of which 2 are ULIRGs and 1 a QSO.
The primary goals of the project are to:
(i) determine the extent and history of star formation, and its relation to
the AGN and torus;
(ii) measure the properties of the molecular gas, and understand its
relation to the torus;
(iii) derive black hole masses from spatially resolved stellar kinematics.
Detailed studies of several individual objects are already published: 
Mkn\,231 \citep{dav:dav04a}; 
NGC\,7469 \citep{dav:dav04b}; 
Circinus \citep{dav:mul06};
NGC\,3227 \citep{dav:dav06a}.
Additionally, we are analysing the general properties of the star
formation (Davies et al., in prep) and the molecular gas (Hicks et
al. in prep).
In this contribution, we highlight 2 pertinent features of the
H$_2$ in NGC\,1068 and summarise our results about nuclear star
formation.

\section{Molecular Gas in NGC\,1068}
\label{dav:sec:n1068}

\begin{figure}
\plotone{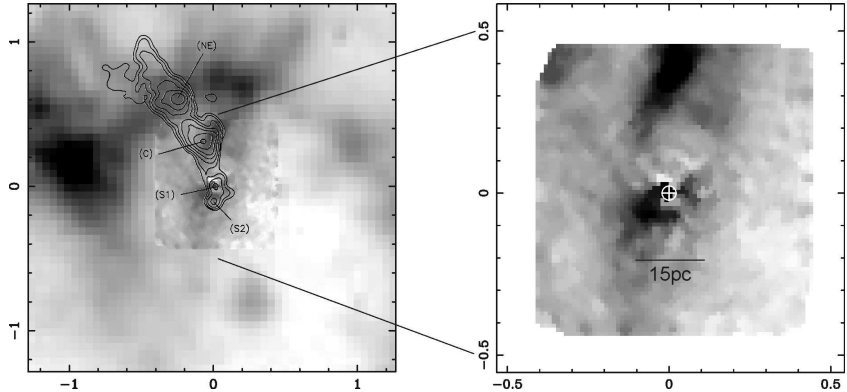}
\caption{H$_2$ 1-0\,S(1) line emission in NGC\,1068.
Left: square-root scaling, with a 5\,GHz radio continuum image
  \citep{dav:gal96} overlaid showing that the jet brightenings
  coincide with the presence of H$_2$;
Right: the central arcsec at a resolution of 85\,mas, showing the
  finger of emission to the north and the central clump of emission
  identified as the torus.}
\label{dav:fig:n1068}
\end{figure}

NGC\,1068 is a prototypical Seyfert~2 galaxy and one
of the cornerstones of AGN unification schemes.
Yet, despite its proximity to us (14.4\,Mpc,
1$^{\prime\prime}=70$\,pc), 
many aspects of its nuclear region remain poorly understood.
In particular, it is now apparent from our SINFONI H$_2$ 1--0\,S(1)
data that simple warped disk models of the molecular gas
\citep{dav:sch00,dav:bak00} cannot account for the fantastic variety
and detail in the morphological and kinematical structure
\citep[see][]{dav:dav06b}.
Here we focus on the two aspects apparent in
Fig.~\ref{dav:fig:n1068}, which are discussed in more detail by 
Mueller S\'anchez et al. (in prep).

\subsection{Jet/Cloud Interaction}

Radio Continuum imaging at 5\,GHz with a resolution of 0.065\arcsec\
revealed a number of structures along the inner part of the radio jet
\citep{dav:gal96}.
Associating component S1 with the inner edge of the torus, these
authors developed a scenario in which component C arises from a shock
interaction between the jet and a dense molecular cloud.
Supporting this hypothesis was the bend in the radio jet, the slightly
flatter spectral index, and the presence of maser emission.
The left-hand panel in Fig.~\ref{dav:fig:n1068} shows that component C
does in fact coincide spatially with a finger of 1-0\,S(1)
emission which traces H$_2$.
Moreover, the $v=1$ levels are thermalised, indicating that the gas is
likely to be rather dense ($>10^4$\,cm$^{-3}$).
Thus the SINFONI data provide direct evidence for the molecular cloud
and hence strongly support the jet-cloud interaction hypothesis.
Interestingly, the jet component NE occurs where the jet crosses an
arc of brighter 1-0\,S(1) emission, suggesting that it may also result
from an interaction between the jet and molecular gas although perhaps
not in a head-on collision as appears to be the case at component C.

\subsection{The Torus}

The right-hand panel of Fig.~\ref{dav:fig:n1068} reveals that close
around the position of the near infrared non-stellar continuum, which
indicates the location of the AGN, we have detected an
extended clump of 1-0\,S(1).
Since the 1-0\,S(1) directly traces H$_2$, we associate this clump
with the molecular material responsible for obscuring the AGN.
There are 2 crucial pieces of evidence to support this:
(i) the 1-0\,S(1) emission is oriented at a position angle of
$\sim120^\circ$, consistent with that of the line
of maser spots \citep{dav:gre96}, the 20\,mas scale radio
continuum in the nuclear component S1 \citep{dav:gal04},
and the 300\,K dust emission (Jaffe, this proceedings);
(ii) the size scale is remarkably similar to those of static torus
models, in particular the more realistic clumpy model of
\cite{dav:hoe06}, for which a size of $15\times7$\,pc (diameter) is
predicted for the H$_2$ distribution (H\"onig, this proceedings).
This then appears to be the first direct image of the torus in
NGC\,1068.

\section{Nuclear Star Formation in AGN}
\label{dav:sec:stars}

In evolving stellar populations, once late
type stars appear, the equivalent widths of the CO\,2-0 2.29\,$\mu$m
and CO\,6-3 1.62\,$\mu$m bandheads stay approximately independent of
age.
Thus, while these features can say nothing about the stellar
population itself, they facilitate the important step of separating the
stellar continuum from the non-stellar continuum associated with the AGN.
They also enable one to measure the kinematics of the stars --
by convolving a suitable stellar template with a
broadening function, which is optimised so as to minimise the
difference between the galaxy spectrum and the convolution product.
Without any additional modelling of the stellar age or history, these
simple measures already yield significant insights into the nuclear
stellar population.

\subsection{Spatially Resolved Nuclear Disks}

NGC\,1068 and NGC\,1097 show the clearest evidence for distinct
nuclear stellar populations from both the surface brightness profile
and the kinematics.
At larger radii (i.e. out to a few arcsec), the surface brightness
profile can be well fit by a single $r^{1/4}$ profile.
But when this is convolved with the PSF and extrapolated inwards, one
finds excess stellar continuum at radii less than 0.5--1\arcsec.
Similarly, at larger radii, the velocity dispersion is
120--150\,km\,s$^{-1}$; but as one approaches the nucleus it
decreases, reaching 70--100\,km\,s$^{-1}$ at the centre.
Sigma-drops have been seen in several spiral galaxies, and are
interpreted as arising from gas accretion into the central regions
followed by star formation \citep{dav:ems06}.
Because the gas is dynamically cool, so will the newly formed stars be:
in contrast to the spheroidal bulge, their distribution will be
rather disky.
We have now spatially resolved 2 of these sigma-drops, and shown that
they are associated with excess stellar continuum -- strong evidence
in favour of their being a distinct dynamically cool stellar
population.
In these specific cases, we are able to trace this population to radii
of $\sim50$\,pc, and estimate a mass of order $10^8$\,M$_\odot$.
Under the assumption that they are self-gravitating, this implies a
vertical scale height of 5--10\,pc suggesting that these nuclear disks
are in fact relatively thick and that random motions still provide
significant support.

\subsection{Luminous Young Starbursts}

Making a rough estimate of the bolometric luminosity L$_{\rm bol}$
from the K-band luminosity L$_K$ is possible without knowing in
detail the star formation history.
This is because L$_{\rm bol}$/L$_K\sim60$ to within a factor 3 (for ages
exceeding 10\,Myr).
Detailed modelling of the stellar populations has in fact been
performed for the AGN listed in Section~\ref{dav:sec:intro}
More generally, making careful corrections for contributions from the AGN
and its associated phenomena, it is possible to constrain the age
using the Br$\gamma$ flux, the supernova rate (estimated from radio
continuum measurements in the literature), the K-band stellar
luminosity, and the dynamical mass.
Doing so, we find for our sample of AGN characteristic ages in the
range 10--300\,Myr.

Looking at how L$_{\rm bol}$ varies as a function of
radius, we find that 
our sample of AGN all follow a similar trend which we are able to trace
from scales of 1\,kpc down to only a few parsecs.
The surface brightness increases at smaller radii, approaching
$10^{13}$\,L$_\odot$\,kpc$^{-2}$ on scales of a few parsec.
This is the surface brightness predicted by models of optically thick
star forming disks in ULIRGs by \cite{dav:tho05}.
The difference is that in ULIRGs it extends over a size scale of
1\,kpc; in these AGN the most intense starburst is confined to the
central few parsec.

\cite{dav:tho05} argued that ULIRGs are essentially Eddington limited
 starbursts. 
The reason is that their bolometric luminosity per unit mass is
 similar to the 500\,L$_\odot$/M$_\odot$ which \cite{dav:sco03} argued
 is sufficient for radiation pressure to halt further accretion.
Within the central few tens of parsecs, the AGN we have observed are
an order of magnitude below this limit.
Intriguingly, the typically low Br$\gamma$ fluxes imply that although the
star formation is recent, it is no longer active.
This is important because short-lived starbursts fade very quickly:
the bolometric luminosity L$_{\rm bol}$ of a burst which is active for
a timescale of 10\,Myr will have decreased by more than an order of
magnitude at an age of 100\,Myr.
Thus in the recent past, these nuclear stellar populations could
 easily have been 10 times more luminous than at present -- in which case
they would have been radiating at the Eddington limit for starbursts.

Our analysis indicates that intense but short-lived starbursts are
likely to be common close around AGN. 
It may be that whether one is able to detect the signature of such
starbursts in AGN depends on the time-scales and size-scales which are
being probed by the observations.



\end{document}